\pdfoutput=1
\documentclass[10pt,aps,prb,amsmath,amssymb,twocolumn,letterpaper,%
         footinbib,showpacs,raggedbottom,balancelastpage,
		citeautoscript,floatfix,reprint]{revtex4-1}

\usepackage{graphicx}
\usepackage[bookmarks=false,colorlinks]{hyperref}
\usepackage[protrusion=true,expansion=true,final]{microtype}

\begin{document}
\title{Normal Mode Determination of Perovskite Crystal Structures with Octahedral Rotations: Theory and Applications}
\author{Mohammad A. Islam}
\author{James M.\ Rondinelli}
\author{Jonathan E.\ Spanier}
\email[ ]{email: spanier@drexel.edu}
\affiliation{Department of Materials Science \& Engineering,\!
	Drexel University,\! Philadelphia,\! PA 19104,\! USA}%
\date{\today}

\begin{abstract}
\noindent Nuclear site analysis methods are used to enumerate the
normal modes of $ABX_{3}$ perovskite polymorphs with
octahedral rotations.
We provide the modes of the
fourteen subgroups of the cubic aristotype
describing the Glazer octahedral tilt patterns, which are obtained from rotations
of the $BX_{6}$ octahedra with different sense and amplitude about
high symmetry axes.
We tabulate all normal modes of each tilt system and specify the
contribution of each atomic species to the mode displacement pattern,
elucidating the physical meaning of the symmetry unique modes.
We have systematically generated 705 schematic atomic displacement
patterns for the normal modes of all 15 (14 rotated + 1 unrotated) Glazer tilt systems.
We show through some illustrative examples how to use these tables to
identify the octahedral rotations, symmetric breathing, and first-order Jahn-Teller
anti-symmetric breathing distortions of the $BX_{6}$ octahedra,
and the associated Raman selection rules.
We anticipate that these tables and schematics will be useful in
understanding the lattice dynamics of bulk perovskites and would
serve as reference point in elucidating the atomic origin of a wide range
of physical properties in synthetic perovskite thin films and superlattices.
\end{abstract}


\maketitle
\sloppy
\section{Introduction}
\begin{figure}
  \centering
  \includegraphics[width=0.49\textwidth,clip]{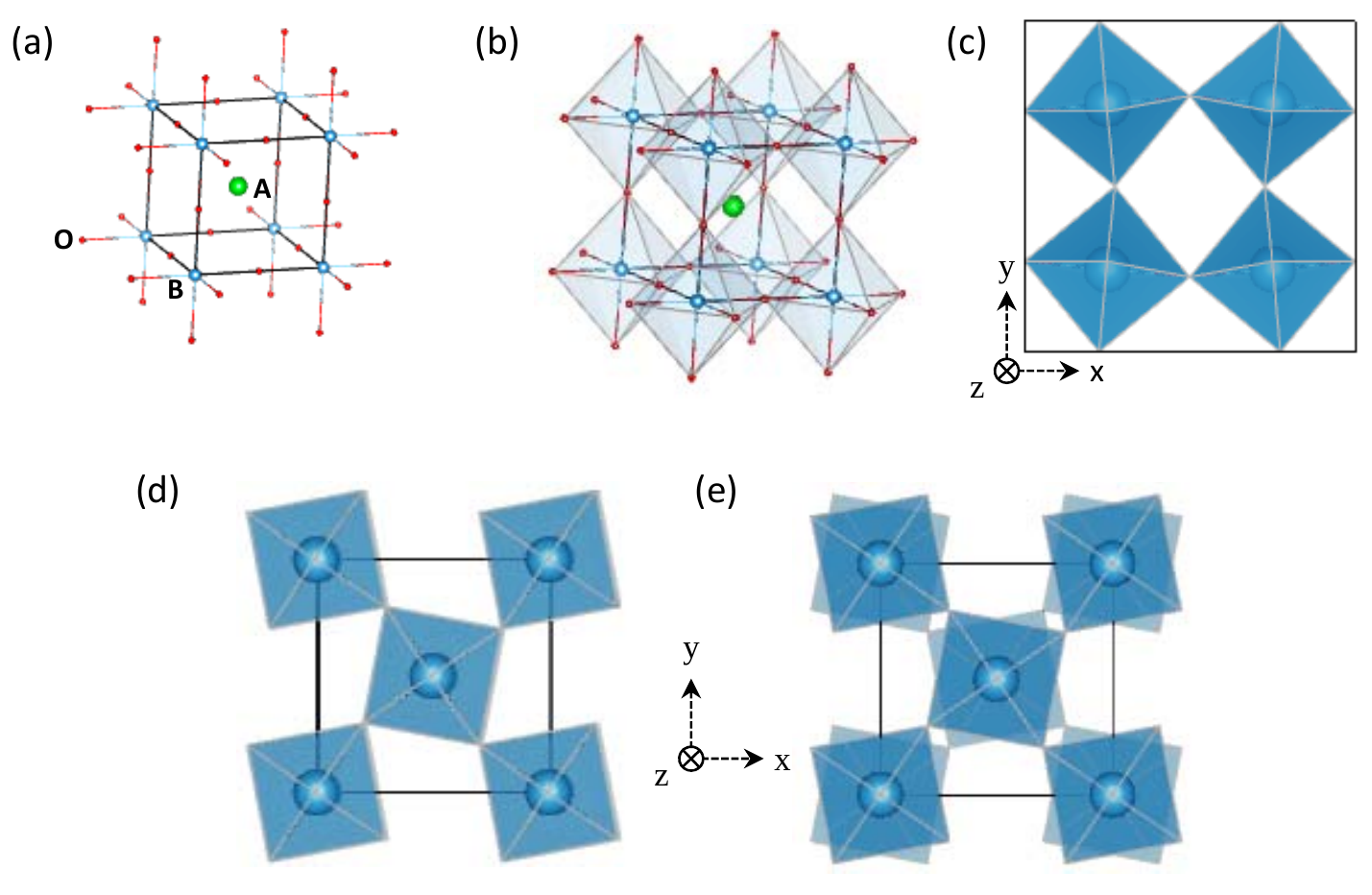}
  \caption{\label{fig:rots} (Color online) The crystal structure of perovskite oxides with the $AB$O$_{3}$ stoichiometry (a)
  The same structure as in (a) but showing the $B$O$_{6}$ octahedral network (b).
Schematic illustration of the the $a^{0}b^{+}b^{+}$ structure (c), indicating that adjacent octahedra
within the same plane rotate in opposite sense to maintain corner connectivity.
Schematic illustration of the
(d) $a^{0}a^{0}c^{+}$  and
(e) $a^{0}a^{0}c^{-}$ rotation patterns.
The in-phase rotations and out-of-phase rotations of the octahedra about
the $z$-axis are discernible in (d) and (e), respectively.
%
%
[Note that for clarity the A-site cations and the oxygen
atoms are omitted from panels (c), (d) and (e)].}
\end{figure}

The compounds with the general formula $ABX_{3}$ ($X\!=\!\textrm{O}$ or a halogen)
where $A$ is either an (alkaline) earth or a rare earth metal and $B$ is a transition metal
with partly filled $d$-orbitals form the perovskite crystal family.
Perovskites are a truly versatile material class that is malleable both
chemically and structurally.
At least 30\% of the elements from the periodic table can occupy the $A$ cation
position and over half of the periodic table fills the $B$ cation position, with
almost 100\% substitution \cite{Schlom:2008}.
The structural adaptability of the $ABX_{3}$ perovskites,
enabled by the flexible and corner-connected $B$X$_{6}$ octahedral network,
explains why such diverse chemistries are compatible in the perovskite crystal structure
(Fig.~\ref{fig:rots}).
The chemical and structural compatibility of oxide-based perovskites
at the atomic-scale
makes it possible to tune their macroscopic properties in a variety of ways, e.g., by judicious
choice of the cations (chemical pressure)~\cite{Woodward1:1997, Woodward2:1997} and hydrostatic pressure~\cite{Medarde:1995}.
Consider the perovskite oxide series $A$NiO$_{3}$, where $A$ is a
rare-earth element: The Ni-O-Ni bond
angles between neighboring octahedra can be made to approach
180$^{\circ}$ (cubic symmetry), driving an insulator--metal
transition as the $A$-site ionic radius  is increased systematically
by substitution of  Eu, Nd, Pr and La \cite{Zhou:2000}.
A modern technique to tune structure-derived electronic properties
in perovskites includes \emph{octahedral engineering}, where
substrate-imposed strain 
and superlattice formation 
are used to alter the magnitude and flavor of the
$B$O$_{6}$ octahedral rotation patterns by exploiting
both misfit strain--octahedral rotation and heterointerfacial
coupling between two similar or dissimilar tilt patterns possessed by the substrate and
film.\cite{Rondinelli:2012, Triscone:2011}
%
The directed control of the crystal structure in epitaxial films
often stabilizes states which are not found in bulk equilibrium phases,
producing dramatic changes in their macroscopic properties.
Control over macroscopic properties requires an understanding of the
atomic structure and an appropriately facile technique to measure
both the property-dictating structural units -- in this case the $B$O$_6$ octahedra --
and their response to external stimuli.
In other words, appropriate experimental methods are required to determine
the structural transitions in functional
perovskite oxides induced by hydrostatic or chemical pressures, strain and superlattice formation.
However, the measurement of oxygen positions in ultrathin $AB$O$_{3}$ films,
required to identify the octahedral tilt patterns, is non-trivial for a variety of
experimental reasons:~\cite{Steve:2010}
Thin films present limited sample volume and often are grown on relatively
thick substrates, which in addition to weak scattering from the oxygen atoms,
makes structure determination of the film challenging.
Raman spectroscopies, along with real and reciprocal space imaging~\cite{Stemmer:2012, ORNL:2008}, however, are experimental
techniques that can reliably determine octahedral tilt patterns.
Raman scattering by zone-center, $k=(0,0,0)$, phonons is an established method for identifying
structural phases in solid, thin-film,
and nanostructured semiconductor and oxide materials, since vibrational modes provide a
unique signature of the crystal structure.
Analyses of the appearance, energies, relative intensities, linewidths and lineshapes, and
symmetry of Raman peaks collected have shown Raman scattering to
be indispensable for investigating important relationships among
correlated structural, electronic charge and spin, and dipolar degrees of
freedom in complex oxide materials \cite{Tenne:2008}.
Despite the optical diffraction-limited spatial resolution, Raman scattering
can probe the onset of a new  phase with local correlation lengths of a
few unit cells.
Furthermore, judicious use of the  optical polarization vectors' relationship with
respect to the plane of the sample, which is obtained through cross sectional
sample preparation, can selectively enhance the Raman scattering intensity
by selected phonons, particularly in the vicinity of a phase phase transition.
Raman spectroscopy studies on cubic $AB$O$_{3}$ type perovskite systems (e.g., BaTiO$_{3}$)
have been carried out by DiDomenico \emph{et al.} and Scott, focusing on the
the splitting of the Raman active modes due to the cubic-to-tetragonal
transformation in these systems~\cite{DiDomenico:1968, Scott:1969}.
More recently, Raman spectroscopy on $AB$O$_{3}$ type
perovskite oxides have focused on the orthorhombic and the
rhombohedral structures~\cite{Abrashev:1999, Ghosh:2005}.
The primary displacement patterns investigated in these studies include
($i$) the soft modes associated
with the rotation of the rigid $B$O$_{6}$  octahedra about different
crystallographic axes, and ($ii$) the higher energy
symmetric and anti-symmetric octahedral breathing-type
modes associated with first-order Jahn-Teller effects.

Dubroka \textit{et al.} have shown in 
La$_{1-y}$Sr$_{y}$Mn$_{1-x}$$M$$_{x}$FeO$_{3}$
($M\!=\!\textrm{Cr, Co, Cu, Zn, Sc or Ga}$)
that the frequency of the soft mode evolves linearly
with the octahedral rotation angle
about the [110]-pseudocubic direction with a positive slope
of 20 cm$^{-1}$ per degree--rotation~\cite{Dubroka:2006}.
Since the oxygen positions dictate the octahedral tilt angle,
Raman scattering provides a non-destructive and readily accessible
means to determine the oxygen positions in $AB$O$_{3}$ perovskites,
for which, alternative experimental methods require substantial
investments in sample preparation, equipment and/or data processing time.\\
\indent Higher-frequency modes -- the Jahn-Teller activated
antisymmetric breathing modes and the
symmetric breathing modes -- often soften across \emph{electronic}
transitions.
For this reason, monitoring the evolution in
Raman modes associated with breathing-type octahedral
distortions can shed light on the origin and onset of charge ordering metal--insulator transitions in
bulk perovskites or superlattices formed by interleaving such materials~\cite{Iliev:2001}.
In some orthorhombic structures, the Jahn-Teller activated modes are
compatible with the crystal structure, i.e., there is no symmetry change,
due to two different types of $B$--O bond lengths induced by the mode.
In contrast, the identical $B$--O bonds present in
rhombohedral structures would require that the Jahn-Teller activated
modes reduce the crystal symmetry---the appearance of these modes in a
Raman spectrum would reveal a symmetry-breaking phase transition.
The combined sensitivity of Raman spectroscopy to symmetry,
structural lattice dynamics and correlated electronic degrees
of freedom, and versatility (e.g., facile application of fields and temperature)
makes the technique well-suited for probing the capabilities of newly identified
octahedral engineering approaches \cite{Rondinelli:2012, Steve:2010}.
Reports on Raman scattering selection rules and experimental work on
perovskite structures realized through octahedral engineering, however,  are rare.
In this work we present a comprehensive list
of the normal modes, their vector displacement patterns, and the associated selection rules
for the 15 octahedral rotation patterns of the $AB$O$_{3}$ perovskite system (Glazer systems).
We have systematically generated 705 schematic atomic displacement
patterns for the normal modes of all 15 (14 rotated + 1 unrotated) Glazer tilt systems.
Furthermore, we have shown how some recent experimental findings can be analyzed using our results and the schematic
diagrams to draw decisive conclusions.
Our results should be applicable immediately to the
complete range of bulk $AB$O$_{3}$ perovskites with octahedral tilting.
We further anticipate that
equipped with comprehensive knowledge of the parent
materials one should be able to analyze and rationally
design more complex structures (e.g., epitaxial films and superlattices)
composed of these $AB$O$_{3}$ perovskites.

%
\section{Structural Distortions in Perovskites}
The  $AB$O$_{3}$ type perovskite is built from basic building units of
$B$O$_6$ octahedra that are corner-connected with additional charge balancing
$A$-site cations occupying interstitial positions
(Fig.~\ref{fig:rots}).
The ideal, aristotype structure, is simple cubic with space group
$O_{h}^{1}$  or equivalently $Pm\bar{3}m$ in the
Hermann-Mauguin notation.
The $A$ and $B$ cations occupy sites with full cubic symmetry
(point group $O_{h}$), whereas the O anions occupy sites with $D_{4h}$
symmetry.
In the ideal structure the ionic radii $r_{A}$, $r_{B}$ and $r_\textrm{O}$
are geometrically related by a characteristic tolerance factor
$t = (r_{A} + r_{O}$)/$\sqrt{2}(r_{B} +r_{O}$) = 1 \cite{Wolfram:2006}.
The majority of the perovskites, however, are not cubic.
The ionic radii of $A$, $B$ and O rarely satisfy the condition that $t=1$.
As a result, the ideal cubic structure undergoes distortions: Typical distortions
are rotations of the $B$O$_{6}$ octahedra about one or more
high-symmetry axes, (anti)-parallel displacement of the
$A$ or $B$ cations away from their cubic site
symmetry, or some combination thereof.\\
\indent The $AB$O$_{3}$ type perovskites can also deviate from the ideal cubic structure by
distortions to the rigid  $B$O$_{6}$ octahedra, i.e., by changes in $B$--O bonds lengths.
These types of distortions are induced by bond valence requirements, orbital degeneracies,
polar distortions and/or valence fluctuations.
Changes in $B$--O bond lengths, for example, are possible and expected
in systems with $d^{4}$ or $d^{7}$ electronic configurations on the
B atoms, leading to Jahn-Teller type antisymmetric breathing of the $B$O$_{6}$ octahedra.\\
\indent The issue of the distortion of the octahedra driven by steric constraints requires special discussion. In the
aristotype $Pm\bar{3}m$, the octahedra are necessarily regular.
The other polymorphs allow distortions and, in general, occur in real
systems.  Stokes \emph{et al.} showed that under certain conditions, however, the geometry
may not require them~\cite{Stokes:1998}.  Consider the simple tilt system $a^{0}a^{0}c^{+}$ (No. 127, \emph{P}4/\emph{mbm}); this system permits octahedral
distortions, but does not necessitate them. If the octahedra are rotated by angle $\phi$ around the \emph{z} axis, then it
can be shown that the octahedra will be regular provided $a/c = \sqrt{2}\cos \phi$.\\
\indent If the ratio of {a/c} differs from this
(and in general it will), the octahedra will be either axially elongated or compressed. Nonetheless,
the bond length distortions are almost always compatible with the symmetry reduction induced by the
antiferrodistortive rotations.
Indeed, Stokes \emph{et al.} examined the geometrical constrains for the octahedral
distortions and found that among the 15 perovskite systems with simple rotation patterns,
only one tilt pattern (space group \emph{P}4$_{2}$/\emph{nmc}, no. 137) requires
distortions to the octahedra, i.e., it cannot be accommodated without changes in the $B$--O bond lengths~\cite{Stokes:1998}.
In general, departure from the cubic symmetry to a structure of lower
symmetry will occur with both rigid rotations of octahedra and bond length distortions as the latter,
secondary modes, often produce further energy stabilization of the crystal structure.

\subsection{Octahedral Rotation Syntax}
\indent The classification of perovskite crystals according to the
rotation patterns of rigid $B$O$_{6}$ octahedra is attributed to
Glazer \cite{Woodward1:1997, Woodward2:1997, Stokes:1998, Glazer:1972, Glazer:1975, Glazer:2011}.
The corner-connectivity of the $B$O$_{6}$ octahedra  in the perovskite structure
constrains the rotations of the adjacent octahedra.
The rotations of an octahedron about a given axis requires that
successive octahedra along directions perpendicular to that axis rotate in the opposite sense about that axis.
Fig.~\ref{fig:rots}(c) shows that a positive rotation (e.g., counterclockwise) of an octahedron about the
$\hat{z}$-axis with magnitude $a$ results in a negative (clockwise)
rotation about $\hat{z}$ of equal magnitude of the adjacent octahedra located
along the $\hat{x}$ and $\hat{y}$ directions.
However, successive octahedra along the rotation axis, $\hat{z}$,
can have either the same or opposite rotation sense [Fig.~\ref{fig:rots}(d) and 1(e)].
To simplify and generalize this description, Glazer introduced a notation whereby
rotations in perovskites are classified according to how adjacent octahedra rotate
along a particular Cartesian axis passing through the $B$-site: $a^{-}$($a^{+}$)
indicates out-of-phase (in-phase)
rotations while $a^{0}$ denotes no rotations.
For example, in a perovskite with the rotation pattern $a^{+}b^{-}b^{-}$ the adjacent
octahedra along the $x$-axis rotate in phase (e.g., all clockwise) and the adjacent
octahedra along the $y$ and $z$ axes rotate out of phase by the same angle $b$.
Examples of the two single tilt systems, $a^{0}a^{0}c^{+}$ and $a^{0}a^{0}c^{-}$, are shown
schematically in Fig.~\ref{fig:rots}(d) and 1(e).
Rotations of the octahedra double the repeat distances perpendicular to the rotation
axis and can result in unit cell vectors inclined to each other by angles other than
90$^{\circ}$ due to ferroelastic strains \cite{Salje:1976}.
The transition into the distorted phase changes the number of formula units in
the unit cell and the full space group symmetry of the crystal system.
By considering all possible rotation senses and amplitudes along various Cartesian
directions, Glazer determined that there are a total of 23 different possible
rotation patterns. (They are also called tilt systems in the literature.)
Some of the tilt systems yield identical space group symmetries:
Group-theoretical analysis by Howard and Stokes \cite{Stokes:1998} shows that of the 23 tilt
systems, 15 simple rotation patterns produce 15 centrosymmetric space
groups.
Table I shows these 15 tilt systems with the appropriate Glazer symbols,
the space group notations (and
numbers in parentheses), the number of formula units (\emph{Z}),
the point symmetry, and known example compounds.
The eight tilt systems that are absent from the group theoretical analysis
were found to have higher symmetry than the corresponding space groups.
Exhaustive studies by Woodward show that 12 of the 15 tilt systems
reported by Howard and Stokes naturally exist(Table I).
None of the eight Glazer tilt systems omitted by Howard and Stokes
have been experimentally observed~\cite{Woodward2:1997}.

\begin{table*}[t]
\begin{ruledtabular}
\caption{\label{tab:lmo_positions}The Glazer systems found in perovskites. The space group symmetries, point symmetries,
formula units, and
experimental compounds, where available, are enumerated for each rotation pattern.  The references for the
experimental compounds can be found in reference 3 and are not be repeated here.``-'' indicates no known example.}
\begin{tabular}{lccc}
Tilt  	 & Space  & Point  & Known  \\
Systems  & Group  &  Symmetry &  Example\\
\hline
1. $a^{0}a^{0}a^{0}$   & $Pm\bar{3}m$ (221) Z = 1	&$m\bar{3}m$ ($O_{h}$) &SrTiO$_{3}$	\\
\hline
2. $a^{0}a^{0}c^{+}$   & $P4/mbm$ (127) Z = 2	&$4/mmm$ ($D_{4h}$) &CsSnI$_{3}$	\\
     & & & (351 - 425 K)\\
\hline
3. $a^{0}b^{+}b^{+}$   & $I4/mmm$ (139) Z = 8	&$4/mmm$ ($D_{4h}$) & - 	\\
 \hline
4. $a^{+}a^{+}a^{+}$   & $Im\bar{3}$ (204) Z = 8	&$m\bar{3}$ ($T_{h}$) &Ca$_{0.25}$Cu$_{0.75}$MnO$_{3}$	\\
\hline
5. $a^{+}b^{+}c^{+}$   & $I/mmm$ (71) Z = 8	&$mmm$ ($D_{2h}$) & - 	\\
  \hline
6. $a^{0}a^{0}c^{-}$   & $I4/mcm$ (140) Z = 4	&$4/mmm$ ($D_{4h}$) &SrTiO$_{3}$	\\
& & & ($<$110 K)\\
\hline
7. $a^{0}b^{-}b^{-}$   & $Imma$ (74) Z = 4	&$mmm$ ($D_{2h}$) &PrAlO$_{3}$	\\
& & & ($~$151 - 205 K)\\
\hline
8. $a^{-}a^{-}a^{-}$   & $R\bar{3}c$ (167) Z = 4	&$\bar{3}m$ ($D_{3h}$) &LaNiO$_{3}$ 	\\
\hline
9. $a^{0}b^{-}c^{-}$   & $C2/m$ (12) Z = 4	&$2/m$ ($C_{2h}$) &PrAlO$_{3}$	\\
& & & ($<$135 K)\\
\hline
10. $a^{-}b^{-}b^{-}$   & $C2/c$ (15) Z = 4	&$2/m$ ($C_{2h}$) &- 	\\
\hline
11. $a^{-}b^{-}c^{-}$   & $P\bar{1}$ (2) Z = 2	&$\bar{1}$ ($C_{i}$) &WO$_{3}$	\\
& & & ($~$230 - 300 K)\\
\hline
12. $a^{0}b^{+}c^{-}$   & $Cmcm$ (63) Z = 8	&$mmm$ ($D_{2h}$) &SrZrAlO$_{3}$	\\
& & & ($~$973 - 1103 K)\\
\hline
13. $a^{+}b^{-}b^{-}$   & $Pnma$ (62) Z = 4	&$mmm$ ($D_{2h}$) &LaMnO$_{3}$	\\
\hline
14. $a^{+}b^{-}c^{-}$   & $P2_{1}/m$ (11) Z = 4	&$2/m$ ($C_{2h}$) &GaLiBr$_{3}$ 	\\
\hline
15. $a^{+}a^{+}c^{-}$   & $P4_{2}/nmc$ (137) Z = 8	&$4/mmm$ ($D_{2h}$) &CaFeTi$_{2}$O$_{3}$\\
\end{tabular}
\end{ruledtabular}
\end{table*}

\section{Crystal Symmetry: Point Groups and Space Groups}
Before describing the method for normal mode and selection rule determination in
the Glazer systems, we provide a brief discussion of point groups, space groups and the
relevant symmetry issues pertinent to this analysis.
\emph{Crystallographic Point Groups}.---%
The description of a physical crystal requires describing the underlying Bravais lattice - an infinite array of discrete
points that appears identical from every point of the array -  and the arrangement of atoms or molecular building
blocks within a primitive cell.
Each crystal system is characterized by a set of rigid
operations that takes the system onto itself. This set of operations is known as the symmetry
group of the crystal system and includes all translations through the lattice vectors and the
point symmetry operations, which are rotations, reflections, and inversions, or some combinations of
these.
Excluding the translations one obtains the subgroup of operations known as the point symmetry
group. Although in principle objects can have many configurations and infinite number of point symmetry
groups, the crystalline environment puts severe limitations on the allowed configurations and results
in 32 crystallographic point groups.

\emph{Space Groups}.---%
When a crystal is formed by inserting a motif into a Bravais lattice, the full symmetry of the crystal depends on
both the symmetry of that object and that of the Bravais lattice; the full symmetry group is called the space
group of the crystal. As a first order approximation, then, the total number of space groups can be simply
compounded out of the 32 crystallographic point groups and the translation symmetry of the motif. There are additional
considerations, however, in the case of the space groups that do not result from simple compounding of the
Bravais lattice systems with the respective point groups.
In the space group, a point is considered invariant
if it is either left in place by a particular symmetry operation or is carried over to a position in an
adjacent unit cell, which is reached by a simple translation of one unit cell.
Also, the presence
of different atomic motifs within the crystal and their different sets of symmetry operations mean that the
rotational axes and the symmetry planes need not coincide at a common point.
As a result more than one set
of operations may fall within the same class. For these reasons, two additional types of compound symmetry
operations appear in space groups: the screw axis and the glide planes are unique for the space groups.
A screw axis operation produces a rotation followed by a translation along the axial direction while a glide
plane operation yields reflection across a plane followed by translation along that plane.
This results in total of 230 space groups. The Glazer systems for perovskites with
octahedral rotations are found in 15  of the general 230 available space groups.

\emph{Site Symmetry}.---%
Within each unit cell defined by a crystallographic space group,
there are special sets of positions in which each
point of the set has identical surroundings, because one or more symmetry elements coincide at that
position.
These points, or sites, and the set of symmetry
operations that leave them invariant
define the \emph{site group} of the set of points; one point belonging
to this set is called a Wyckoff position.
The site group is a subgroup of the space group and is isomorphous with one of the 32 crystallographic
point groups compatible with the space group. The number of equivalent points $n$ in a site group is
equal to the order of the factor group (space group) $H$ divided by the order of the site group $h$, i.e., $n = H/h$.
For example, in the simplest Glazer tilt system without any
octahedral rotations,
$a^{0}a^{0}a^{0}$ (space group $O_{h}^{1}$ or $Pm\bar{3}m$),
the $A$-site atom belongs to the site group $1a$, with
site symmetry $O_{h}$, the $B$-atom belongs to the site
group $1b$  with site symmetry $O_{h}$, and the three oxygen
atoms belong to the site group $3c$ with site symmetry $D_{4h}$.
\footnote {An equivalent setting for the $a^{0}a^{0}a^{0}$
rotation pattern, has the $A$-site at Wyckoff position $1b$, the B-site at $1a$ and the oxygen atom at $3d$.}
The importance of site symmetry becomes
evident in the next section where we discuss the methods of normal mode determination in a crystal.


\section{Normal Mode Determination}
\subsection{General Approaches}
\indent Of the various methods that are available to analyze the primitive unit cell and determine the selection rules for
the first order phonon spectrum \cite{Bhagavantum:1939, DeAngelis:1972, Halford:1946, Mathieu:1945, Porto:1981} the
Nuclear Site Analysis (NSA) method have been found to be most convenient
since it requires the least amount of information about the unit cell; usually the Wyckoff positions and the site groups of the atoms are all that are required.
Since for most practical
purposes the $AB$O$_{3}$ perovskites can be considered as a flexible network of
corner connected $B$O$_{6}$ octahedra, determining the Wyckoff positions and sites
groups are relatively easy for these systems, and we use the NSA method in the present article.\\
\indent To understand the approach in the NSA method, consider for example,
a set of atoms with site group $g$ inside the unit cell of a crystal of space
group $G$.
If the order of $G$ is $H$ and that of $g$ is $h$, then $n = H/h$, where
$n$ is the number of equivalent points (e.g., atoms) in the site
group $g$. Since $g$ is a subgroup of $G$, each irrep of $g$ will map
onto one or more of the irreps of $G$. In the NSA method only the vector irreps of the site group $g$
are chosen and these are then mapped onto the irreps of the crystallographic point group of the space group $G$.
The total number of normal modes of a crystal with a
particular space group can finally be obtained by considering all the site groups
and occupied Wyckoff positions and taking their algebraic sum.

\subsection{Methodology for Perovskites with Octahedral Rotations}

We use the Nuclear Site Group Analysis (NSA) method to enumerate the normal
modes of $AB$O$_{3}$ perovskite oxides characterized by
rotations and steric constraint driven distortions of the $B$O$_6$ octahedra.
We begin with the prototypical aristotype cubic polymorph, e.g.,\ room-temperature
SrTiO$_{3}$, with the symmetry $O_{h}^{1}$ ($Pm\bar{3}m$).
To generate the rotationally distorted phases, we then freeze-in each specific
symmetry lowering rotation pattern (Table I, column 1) into the cubic
structure to obtain the space groups in column 2 of Table I.
The rotations lead to a splitting of the oxygen Wyckoff orbits, which
have been tabulated elsewhere~\cite{Knight:2009}.
%

Knowledge of the occupied cation and anion Wyckoff positions
in the standard notations is essential to the NSA method, because they
are directly related to the site symmetry of the atomic species.
The normal modes of each atom are identified by consulting
the International Tables for Crystallography (ITC) and/or
Porto \textit{et al.} \cite {Porto:1981, ITXRC} for the character tables of the site
groups and of the space group.
These elements are required for determining the vector irreps of the site group and
subsequently mapping these to the irreps of the space group.
We repeated this analysis for all atomic sets until the total number of
normal modes for the asymmetric unit was obtained. The data are presented in Tables III - XVII found in the appendix.

\subsection{Example: Normal Mode Determination of the $a^{+}b^{-}b^{-}$ Rotation Pattern}
We now describe this process in detail for the orthorhombic $a^{+}b^{-}b^{-}$ rotation pattern, Glazer tilt system 13 (Table XV),
which LaMnO$_{3}$ is known to exhibit~\cite{Elemans:1971}. %
When the tilt system $a^{+}b^{-}b^{-}$  is applied to the cubic structure the
space group transforms from cubic $Pm\bar{3}m$ to orthorhombic $Pnma$.
The displacement pattern of oxygen atoms also leads to an approximate
$\sqrt{2}\times\sqrt{2}\times2$ increase in the cell dimensions $a$, $b$ and $c$, respectively,
producing a unit cell with four $AB$O$_3$ formula units ($Z=4$), as enumerated in column 2 of Table I.
In Schoenflies notation, the $Pnma$ space group is equivalent to $D_{2h}^{16}$,
making the relevant point symmetry $mmm$ or $D_{2h}$ (Table XV).

\begin{table}[t]
\begin{ruledtabular}
\caption{\label{tab:lmo_positions}Occupied Wyckoff positions and site symmetries for $Pnma$ orthorhombic
(space group 62, $a=5.743$, $b=7.695$, $c=5.537$~\AA) LaMnO$_3$ from
Ref.~\onlinecite{Elemans:1971}.}
\begin{tabular}{lccllll}
	   & Wyckoff & Site & \multicolumn{3}{c}{Position} \\[-0.1em]
\cline{4-6}
Atoms & Orbit & Symmetries & $x$ & $y$ & $z$ \\
\hline
La 		& $4c$	&	$C_{s}^{xz}$ 	&	0.55	& $\frac{1}{4}$ & 0.009	\\
Mn 		& $4a$	&	$C_{i}$	& 0 & 0 & 0 \\
O(1) 	& $4c$	&	$C_{s}^{xz}$	& -0.011 & $\frac{1}{4}$& -0.071 \\
O(2)		& $8d$	& 	$C_{1}$	& 0.309	& 0.039	& 0.225 \\
\end{tabular}
\end{ruledtabular}
\end{table}

After obtaining the lower symmetry structure (\autoref{tab:lmo_positions}), the Wyckoff positions can be read
from the ITC, or by using crystallographic visualization
software such as \textsc{vesta} and \textsc{CrystalMaker} \cite{ITXRC, VESTA}.
Note that in the Wyckoff orbit column,  the numeral preceding the
Wyckoff label indicates that all of those symmetry equivalent atoms
occupy that particular Wyckoff position.
Chemically equivalent atoms can occupy crystallographically
unique Wyckoff positions in crystals.
This becomes apparent in LaMnO$_3$ with
the $a^{+}b^{-}b^{-}$ rotation pattern:
The eight oxygen atoms found in the $ac$ plane
occupy Wyckoff positions $8d$ and possess site symmetry $C_{1}$, whereas the
four oxygen atoms along the $b$ axis occupy Wyckoff positions $4c$ and possess
site symmetry $C_{s}^{xz}$ (\autoref{tab:lmo_positions}).
This ``splitting'' of the Wyckoff orbits of the O atoms is a signature
of the structural transition on going from $Pm\bar{3}m{\rightarrow}Pnma$,
which can be exploited in Raman spectroscopy studies to recognize the
onset of the transition by monitoring the change  in number of peaks
(modes), intensities, etc., with external stimuli.
In the ideal cubic case, however, only one Wyckoff position is necessary to describe all oxygen
atoms in the crystal structure.

For determining the normal modes, consider the four oxygen atoms occupying the Wyckoff position $4c$ and site symmetry $C_{s}$,
the irreps of which can be found in
the ITXRC and/or Porto \textit{et al.} \cite {Porto:1981, ITXRC}. The vector irreps of $C_{s}$ are $A'$ and $A''$.
When these are mapped on to the irreps of the crystallographic point group $D_{2h}$ of $Pnma$, we obtain
the contributions of these four oxygen atoms to the
normal modes of LaMnO$_3$. This is repeated for all the atomic species in the unit cell.
 The first column of Table XV shows all available normal modes for the
tilt system $a^{+}b^{-}b^{-}$ in the absence of any other atomic distortions
(cation displacements or Jahn-Teller distortions of the ligands).
The values in the  main body of the table shows the number of symmetry modes in which the atoms participate.
This type of presentation, instead of simply writing down all the modes, provides physical meaning to the modes.
As shown in Table XV, the 4 La atoms have site-symmetry $C_{s}^{xz}$ and contribute the following modes:
\[
2A_{g}\oplus  A_{u}\oplus B_{1g}\oplus 2B_{1u}\oplus 2B_{2g}\oplus B_{2u}\oplus B_{3g}\oplus 2B_{3u}\, .
\]
The 4 Mn atoms with site-symmetry $C_{i}$ contribute:
\[
3A_{u}\oplus 3B_{1u}\oplus 3B_{2u}\oplus 3B_{3u}\, ,
\]
while the 4 O(1) atoms with site-symmetry $C_{s}^{xz}$ contribute:
\[
2A_{g}\oplus  A_{u}\oplus B_{1g}\oplus 2B_{1u}\oplus 2B_{2g}\oplus B_{2u}\oplus B_{3g}\oplus 2B_{3u}\, ,
\]
and the 8 O(2) atoms with site symmetry $C_{1}$ contribute:
\[
3A_{g}\oplus  3A_{u}\oplus 3B_{1g}\oplus 3B_{1u}\oplus 3B_{2g}\oplus 3B_{2u}\oplus 3B_{3g}\oplus 3B_{3u}\, .
\]

The last two columns in Table XV provide the mode selection rules in
standard notations.
The acoustic modes belong to representations which contain the translation vectors,
symbolized by $T$.
After subtracting out the three acoustic modes
available to three-dimensional crystals,
the remaining modes labeled by $T$ are IR active.
The selection rules for the  Raman modes  are shown separately in the
last column with the polarizability tensor components,
denoted by $\alpha_{ij}$ and the appropriate Cartesian subscripts.
$\alpha_{ij}$ are also known as susceptibility derivative tensors in the literature.
\begin{figure*}
  \centering
  \includegraphics[width=0.8\textwidth,clip]{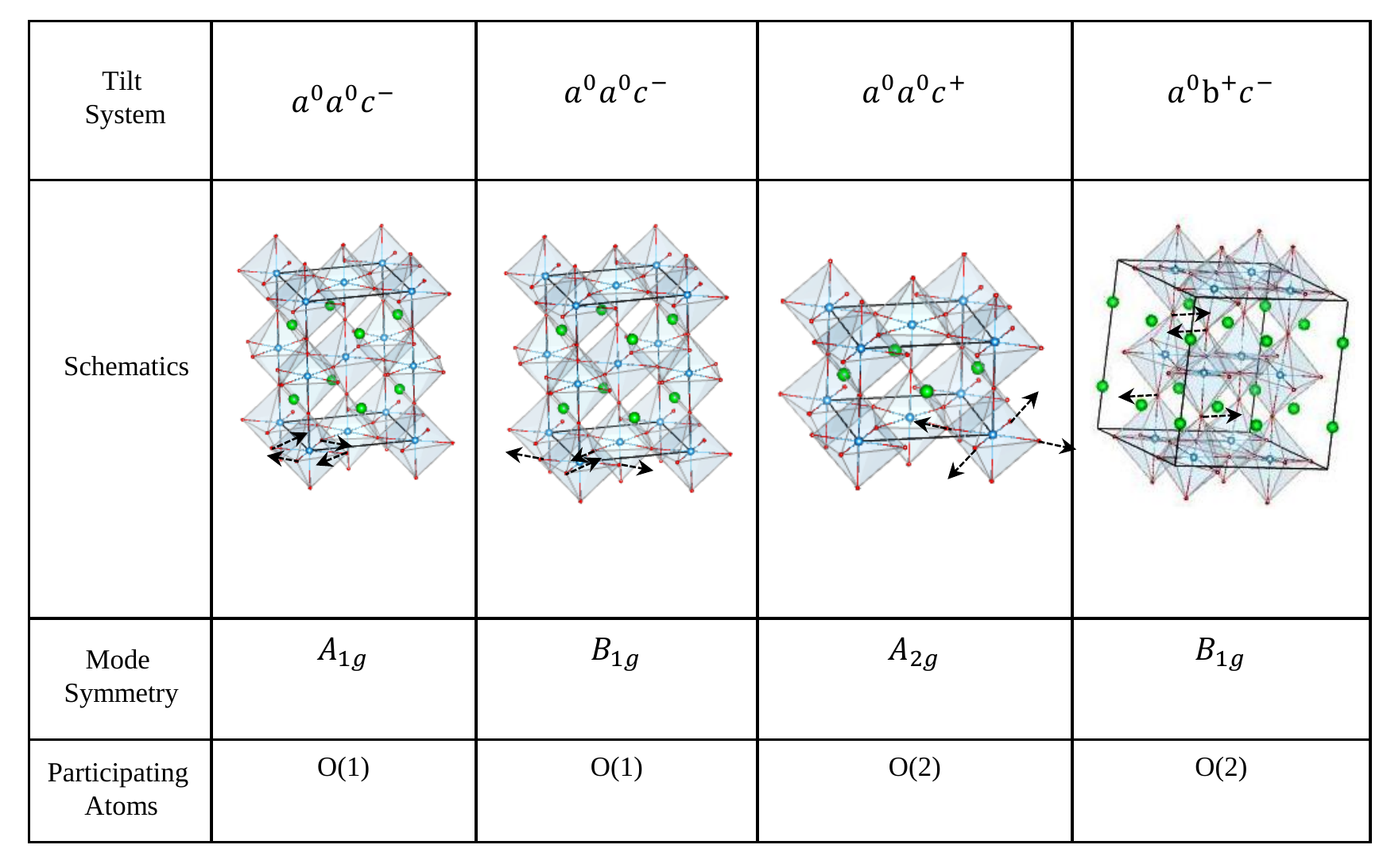}
  \caption{\label{fig:tilt_table}(Color online) Schematic
atomic vibrational patterns of selected normal modes for a few octahedral rotation patterns.
The symmetry mode symbols (irreducible representations) and the atoms participating in the normal modes are also provided.
The complete database for all normal modes in each Glazer tilt
system can be viewed online by accessing: \href{http://meso.materials.drexel.edu/Glazer\_modes}{http://meso.materials.drexel.edu/Glazer\_modes}.
}
\end{figure*}

For $Pnma$ LaMnO$_{3}$, Table XV reveals that one of each of the
$B_{1u}$, $B_{2u}$ and $B_{3u}$ modes account for the
acoustic (translational) modes of the crystal.
The remaining $9B_{1u}$, $3B_{2u}$ and $8B_{3u}$ modes are IR-active, and
the 7$A_{g}$, 5$B_{1g}$, 7$B_{2g}$, and 5$B_{3g}$ modes are Raman active. The normal modes that are neither IR active,
nor Raman active are considered silent.
In the present case, the silent modes transform with $A_{u}$ symmetry.
For completeness, we also provide the rotation (axial)
vectors compatible with the appropriate modes.
For example, as shown in Table XV, the rotation
$R_{z}$ has the same symmetry properties as that of the $B_{1g}$ modes. In general, the
rotations have the same symmetry properties as those
of the difference between anti-symmetric Raman tensors: $R_{z}$ transforms as $\alpha_{xy}  -  \alpha_{yx}$.
For Raman scattering that uses excitation frequency far from
electronic resonance, these differences are zero since the polarizability tensors
are symmetric.
In resonance Raman experiments, however, asymmetries in the polarizability
tensors are introduced and then such modes, i.e., modes with
$R_{x}$, $R_{y}$, $R_{z}$ labels could become Raman active.
The last row of each table shows the total number of modes for each
rotation pattern.
Note that modes with $A$ and $B$ spectroscopic labels are non-degenerate,
whereas modes with $E$ ($T$) labels are doubly (triply) degenerate.

Finally, to complement these tables we have developed a graphical
description of each mode by using the character projection method.
The complete compilation of all 705 of these diagrams
is available by accessing: \href{http://meso.materials.drexel.edu/Glazer\_modes}{http://meso.materials.drexel.edu/Glazer\_modes}.
A few of the representative diagrams are shown in \autoref{fig:tilt_table}.
The vector displacements without such diagrams, corresponding to all normal modes
for each Wyckoff position available to the 230 space
groups, can also be obtained using the Bilbao Crystallographic Server and
the web-based program \href{http://www.cryst.ehu.es/rep/sam.html}{\textsc{sam}}.

\section{Discussions and Applications}
\subsection{Raman Mode Bookkeeping: How many modes are there?}

\indent One way to check the accuracy of the normal mode tables
we have generated for the $AB$O$_{3}$ perovskites with octahedral rotations
(Tables III--XVII) is to count the number of normal modes enumerated for
each crystal system.
If there are $p>2$ different atoms in a three-dimensional \emph{primitive} cell, there should be $3p$
branches to the phonon dispersion relation, and consequently, $3p$ modes at the $\Gamma$ point.
Of these $3p$ modes, 3 are acoustic (one longitudinal and two transverse) modes and the remaining, $3p- 3$, are optical modes.
For space groups with a primitive lattice, as is
 the
case for the Glazer tilt systems 1, 2, 8, 11, 13, 14, and 15,
the Raman mode enumeration is straightforward.
For the Glazer systems with non-primitive lattice, care must be taken to avoid redundancy in
the enumeration of the normal modes. For these systems the usual construction of the primitive
cell is less useful because it doesn't usually reveal the full symmetry of the structure.
In these cases the \emph{crystallographic} cell or the \emph{conventional} cell is constructed by
tiling an integral number of  primitive cells through
lattice translations of the primitive cell lattice point.
%
Since the translation group transforms as the identity element of
the space group, the distribution of degrees of freedom among the space
group irreps does not change.
However, this introduces a redundancy in the value of the
characters for each symmetry: Each set of equipoints
in the primitive cell is duplicated for the additional lattice points
in the expanded crystallographic cell.
Since redefining the unit cell vectors does not affect the rotational
symmetry, the redundancy is uniformly distributed over the
characters of all symmetry classes of the space group.
To recover the primitive (\textsc{p}) cell characters,
one should divide the conventional (\textsc{c}) cell characters by the
lattice point multiplicity, i.e.,
$
\chi_\textrm{\textsc{p}} = \chi_\textrm{\textsc{c}}/\ell\, ,
$
%
where $\ell$ is the number of lattice points (or primitive cells) in the
full crystallographic cell: $\ell = 4$ for face-centered cell, $\ell = 3$
for rhombohedral \emph{R}-centered cells given in hexagonal settings, and $\ell = 2$ for both
body centered cells and base centered cells \cite{DeAngelis:1972}.
This consequently yields the number of normal modes being an integral fraction of 3\emph{q}, where $q$ is the number
of atoms in the conventional cell.
The Glazer systems 3, 4, 5, 6, and 7 are body centered and as such the number of modes is $3q/2$.
The Glazer system 8 is rhombohedral and as such the number of modes is $3q/3$.
The Glazer systems 9, 10 and 12 are base centered and as such the number of modes is $3q/2$.
\subsection{Applications: Phase Transitions in  Bulk Perovskites}
\begin{figure}
\centering
  \includegraphics[width=0.43\textwidth,clip]{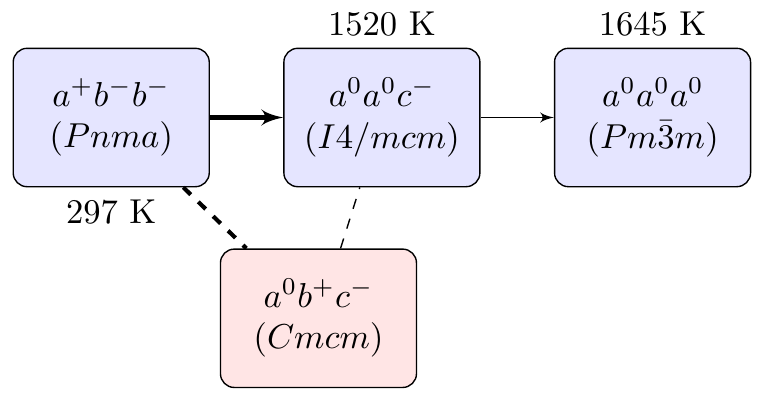}
\caption{\label{fig:cto_pt}
Sequence of structural phase transitions and the change
in octahedral rotation patterns that occurs with increasing temperature
in perovskite CaTiO$_3$. The orthorhombic $Cmcm$ phase (linked by broken
lines) appears (or is sometimes absent) experimentally as an
 intermediate phase between $Pnma$ and $I4/mcm$ at intermediate
temperatures.
Transitions required by Landau theory to be first (second) order are
distinguished by heavy (light) lines.}
\end{figure}

\indent Having ensured that the number of modes for each system
is accurate, we now provide some illustrative applications and
discuss a few limitations
of the constructed tables and the schematics. 
Here we show how the Raman normal modes determination of
the various octahedral rotation patterns in perovskites provides a manner
to explore antiferrodistortive phase transitions.
CaTiO$_{3}$ is a widely used electroceramic and has
been recognized as a suitable material for the immobilization of
high-level radioactive wastes \cite{Ringwood:1988}.
Consequently, the properties of CaTiO$_{3}$, especially its crystal structures
over different temperature regimes have lately been the focus of intense studies.\cite{McMillan:1988}
It undergoes a series of structural
transitions driven by the octahedral rotations (\autoref{fig:cto_pt}):\cite{Yashima:2009}
Under ambient conditions, CaTiO$_{3}$ is orthorhombic
(space group $Pnma$) and exhibits the $a^{+}b^{-}b^{-}$
rotation pattern.
Upon heating, the structure transforms to tetragonal
(space group $I4/mcm$, Glazer system $a^{0}a^{0}c^{-}$)
at 1520 K.
Note that some studies suggest that between the tetragonal and the  orthorhombic phase there is an additional structure with orthorhombic
symmetry,  \emph{Cmcm}, that exhibits one in-phase and one out-of-phase rotation
($a^{0}b^{+}c^{-}$).
Finally near 1645~K, CaTiO$_3$ transforms into the ideal cubic $Pm\bar{3}m$
structure without any octahedral rotations ($a^{0}a^{0}a^{0}$).

Unpolarized Raman studies enable identification of the nature and onset of the phase transition from the
octahedrally rotated structures to the cubic structure upon heating.
The relevant information for the analysis of the rotation-driven structural transitions
\[
a^{+}b^{-}b^{-}
\rightarrow
a^{0}b^{+}c^{-}
\rightarrow
a^{0}a^{0}c^{-}
\rightarrow
a^{0}a^{0}a^{0}
\]
are found in Tables XV, XIV, VIII and III, respectively.
First, comparing these tables reveals that all of the rotationally
distorted perovskite polymorphs have
first-order Raman active vibrational modes, whereas
all zone-center modes in cubic ($a^0a^0a^0$) structure are
Raman inactive.
%
Accordingly, unpolarized Raman spectra of the tilted structures
should exhibit first-order Raman features, possibly superimposed
on broad second order features, where as the cubic should not. 
(Note that in practice, however, at the transition to the cubic
structure the first order Raman features could give way to
broad second order features---such effects
are seen in prototypical cubic perovskite SrTiO$_{3}$~\cite{McMillan:1988}.)

Polarized Raman studies can be used to determine the phase
transitions among the tilted structures. Raman scattering cross sections ($\sigma_R$) for
Stokes scattering is related to the polarizations of the
incident ($\epsilon_{I}^{i}$) and scattered ($\epsilon_{S}^{j}$)
radiations and the susceptibility (derivative) tensors by the following formula:
 \[
 \sigma_R  \propto | e_{S}^{i} e_{I}^{j}\alpha_{ij}|^{2}\, ,
\]
\noindent where $\alpha_{ij}$ are the components of the susceptibility (derivative) tensors appropriate for a given mode,
shown in the last column of the Tables III--XVII.
Note that summation over the repeated cartesian indices ($i$, $j$) is implicit in the above equation.
As an application of this, let's consider
Raman spectroscopy in the back scattering geometry for CaTiO$_3$.
The standard scattering geometry for this experiment is $z(ij)z$, which is shorthand for
$k_{I}(\mathrm{e}_{I}^i,\mathrm{e}_{S}^j)k_{S}$, where $k_{I}$
and $k_{S}$ are the wave vectors of the incident and scattered light,
respectively, and $\mathrm{e}_{I}^i$ and $\mathrm{e}_{S}^j$ are their
corresponding polarization vectors. Since both $k_{S}$ and $k_{I}$ are in the $z$ direction, the
polarization vectors of both the incident and scattered radiation are in the $xy$-plane,
resulting in Raman cross section of:
\[
 \sigma_R\propto |e_{S}^{x} e_{I}^{x}\alpha_{xx}+ e_{S}^{x} e_{I}^{y}\alpha_{xy}+
 e_{S}^{y} e_{I}^{x}\alpha_{yx}+e_{S}^{y} e_{I}^{y}\alpha_{yy}|^{2}\, .
\]
Consider now the transition from $a^+b^-b^-\rightarrow a^0a^0c^-$.
As seen from Tables XV and VIII, there are significant differences between
the mode distributions in the two structures.
In the $a^+b^-b^-$ ($Pnma$) structure, the Raman active modes are:
\[
7A_{g} \oplus 5B_{1g}  \oplus  7B_{2g} \oplus 5B_{3g}\,,
\]
whereas in the $a^0a^0c^-$ (\emph{I}4/\emph{mcm}) system, the Raman active modes are:
\[
A_{g} \oplus B_{g} \oplus 2B_{2g} \oplus E_{g}\, .
\]
Based on the larger number of Raman modes in $Pnma$
(24 compared to the 8 in $I4/mcm$ CaTiO$_3$), there should be a readily
detectable difference in the Raman spectra, i.e.,
one should expect significantly more first-order Raman features in the
$Pnma$ structure than in $I4/mcm$, even in unpolarized Raman spectroscopy.
The two phases can further be distinguished by performing a careful polarized
Raman experiment, whereby an appropriate selection of the incident and
scattered light orientation will enhance or suppress a first-order Raman
mode. Raman spectroscopy in the $z(xx)z$ geometry will result
in Raman cross sections of
\[
 \sigma_R\propto |e_{S}^{x} e_{I}^{x}\alpha_{xx}|^{2}\, .
\]
\noindent This will only show modes of $A_{g}$ symmetry of which are seven in the $Pnma$ and
only one in the $I4/mcm$ structure.
Similar results can be expected in the $z(yy)z$ geometry.
Further careful studies with the $z(yz)y$ configuration would
show first order Raman features from the singly degenerate $B_{3g}$ mode
in $Pnma$ CaTiO$_3$, whereas an $E_{g}$ doublet would appear
in the $I4/mcm$ structure.
The controversial phase transition from $Pnma$ to $Cmcm$
would be more difficult to determine
by examining the Raman spectra alone, since the Raman active
modes present in the two structures are
similar (cf.\ Tables XV and XIV).
To remedy the ambiguity, one could rather use the octahedral
rotation angle, specifically the component about the $x$-axis,
as an order parameter for the $Pnma \rightarrow Cmcm$
transition, because it goes from a finite to null value as
the rotation pattern changes, i.e.\ $a^{+}b^{-}b^{-}$ to $a^{0}b^{+}c^{-}$.
%
%
Recently, Dubroka \emph{et al.} and Abrashev \emph{et al.}
have shown how to correlate the Raman $A_{g}$ mode frequency
with the tilt angle, and subsequently that 
tilt angle with the free oxygen positions~\cite{Dubroka:2006, Abrashev:1999}.
Their results, along with our schematic diagrams for the normal modes of
the rotationally distorted perovskites and lattice dynamic calculations
could be used to ascertain the phase transition between these two structures~\cite{Guo:2011, Yashima:2009}.

\subsection{Applications: Structure--property Relations in
Perovskite Thin Films and Superlattices}
We expect this methodology to have higher impact in $AB$O$_{3}$ perovskite
systems realized via octahedral engineering route whereby  efforts are
being made to exploit epitaxial strains, interfacial octahedral coupling, and superlattices to
stabilize structures inaccessible via conventional solid-state
chemistry techniques.
Recent computational studies predict selective control of the octahedra about
different crystal directions by using epitaxial  strain to control the rotation--lattice
coupling in epitaxial films~\cite{Zayak:2006, Spaldin:2010}.
Strain can therefore be used to directly tailor the flavor -- magnitude and phase --
of octahedral rotation patterns in thin films.
The normal mode determination presented here, could be used to design and
analyze $AB$O$_{3}$ films with octahedral rotations.
Consider for example, La$_{1/3}$Sr$_{2/3}$FeO$_{3}$ (LSFO), which
is rhombohedral (Glazer tilt system $a^{-}a^{-}a^{-}$) in the bulk.
Its normal modes appear in Table X.
However, if an epitaxial film of LSFO is grown on \emph{cubic} SrTiO$_{3}$,
it would likely transform to a monoclinic phase with the $a^{-}b^{-}b^{-}$
octahedral rotation pattern due to lattice mismatch between the film and
the substrate in the epitaxial plane.
Reduction of symmetry will lead to a splitting of some degenerate modes of the
rhombohedral system, which can be determined by using the  standard reduction formulae\cite{Tsukerblat:2006} and our Table XII, i.e.,
the relevant normal modes for the \emph{monoclinic} LSFO.
By comparing the Raman spectra of bulk LSFO and the epitaxial film, it should be
possible to evaluate the change in crystal structure for the rhombohedral to monoclinic
transition.
Formation of a coherent perovskite--perovskite heterointerface produces a
geometrical constraint of the octahedral connectivity across the interface in
synthetic structures \cite{Rondinelli:2012}.
The octahedral rotations present in a substrate, for example, can propagate into
the near-interface region of a thin film:
This was recently observed experimentally using high-resolution x-ray diffraction and
synchrotron radiation, in
(LaNiO$_{3}$)$_{n}$/(SrMnO$_{3}$)$_2$ superlattices~\cite{Steve:2011}.
This so-called \emph{octahedral proximity effect} remains to be harnessed for materials
design, yet has the ability to mediate electron--lattice coupling across interfaces and
produce remarkable changes in the electronic properties of ultrathin films and superlattices.
Segal \emph{et al.} have shown that the propagation of lattice
vibrational modes (phonons) associated with the dynamic octahedral rotations causes
drastic changes in the resistivity of a La$_{0.53}$Sr$_{0.47}$MnO$_{3}$ (LSMO) films as the system is cooled through
the $a^0a^0a^0 \rightarrow a^0a^0c^-$ phase transition of the SrTiO$_{3}$ substrate.
The proposed microscopic mechanism was attributed
to evanescent cross-interface coupling between the charge carriers
in the film and the soft phonon mode responsible for the out-of-phase
rotation of TiO$_6$ in SrTiO$_{3}$.
In other words, enhanced electron--lattice coupling mediated  through the
correlated oxygen octahedral motions~\cite{Segal:2011}.
Other external stimuli besides temperature can promote novel
behavior mediated by octahedral rotations.
For example, Caviglia \emph{et al.} have shown that an optical excitation in
resonance with a stretching mode of a perovskite substrate can
trigger a dynamic (transient) insulator to metal transition in NdNiO$_{3}$
epitaxial films \emph{below} the bulk N\'{e}el temperature~\cite{Caviglia:2012}.
Recent Raman spectroscopy studies on superlattices of
superconducting YBa$_2$Cu$_3$O$_7$ and magnetoresistive
La$_{2/3}$Sr$_{1/3}$MnO$_3$  below the critical temperature of the cuprate, reveal
that the superconducting state can renormalize the lineshape of the MnO$_6$
octahedral rotation modes in the manganite, providing an unexplored platform to tailor
electron--lattice interactions in correlated materials~\cite{Driza:2012}.

The unique effects described here are all provided by the normal mode distributions
within the materials and their coupling to the low-energy electronic structure.
They could possess tremendous technological potential, yet the microscopic
mechanisms in nearly all cases remain inadequately understood and characterized.
Much of the current results of computational studies, for example,
await experimental verifications.
Segal \emph{et al.} suggested that the change in conductivity in their
LSMO epitaxial film could be due to the alteration of the $B$--O--$B$
bond angles in the film by the soft phonon modes of the STO substrate---direct
measurement of these variations remains to be seen.
Note however that they also offered an equally likely scenario:
The substrate could modify the $B$--O bond
lengths in the film, resulting in similar modifications to  the films conductivity, via
a bond stretching mode.
In the experiments by Caviglia \emph{et al.}, the authors acknowledge that the
precise nature of the interaction between the LAO substrate and the
NdNiO$_{3}$ film remains poorly understood and merits systematic
studies.
For that reason, the group offered an alternative explanation for the
dynamic insulator--metal transition:
The stretching mode of the substrate could couple to a rotational mode
(soft mode) of the film, which would also alter the films conductivity through a
modification of the Ni $3d$ -- O $2p$ angular orbital overlap.
For such complex systems, our results presented here, especially the normal mode
schematics, could be useful in drawing more decisive conclusions.
By superimposing a unit cell of an epitaxial material on a unit cell of the intended substrate
and inducing a particular normal mode in the substrate, once could examine geometrically and on symmetry grounds
the extent to which the substrate modes couple to the film and
induce changes in both the topology of the film's perovskite lattice and properties.
Since our paper presents a complete list of normal modes, selection rules, and
schematic diagrams of the bulk $AB$O$_{3}$ perovskites with octahedral
rotations, careful implementation of our results could guide research
avenues in search of quantum phase transitions in artificial materials using optical
excitation of lattice modes.
\subsection{Applications: Normal Mode Determination of a Non-Glazer System}
As an extension of the methodology presented here, we describe the method for normal mode determination of a
�non-Glazer� system that undergoes rotation of the $B$O$_6$ octahedra as well as off-centering displacements of the ions:
BiFeO$_3$ is a magnetoelectric multiferroic whose ground state exhibits both antiferromagnetic and ferroelectric order.
Bulk BiFeO$_3$ has the rhombohedral
\emph{R}3\emph{c} structure, which consists of antiferrodistortive octahedral rotation ($a^{-}a^{-}a^{-}$) around the [111] direction
and an additional relative off-centering of anions and cations along the [111] direction leading to a
ferroelectric polarization around that axis. Using method similar to LaMnO$_3$ (Sec. IV C)
we construct the normal mode table for BiFeO$_3$ in the \emph{bulk} (Table XVIII).\\
\indent These bulk normal modes, however, should change when an epitaxial film of BiFeO$_3$ is grown under epitaxial strain on
a substrate with different symmetry. If rhombohedral BiFeO$_3$
is deposited on a cubic (001) substrate like STO, the film, under biaxial strain, will acquire monoclinic
symmetry due to the geometric constraints imposed by epitaxy. In this case the Glazer notation of this film becomes $a^{-}b^{-}b^{-}$ and the relevant normal mode table to use is Table XII.
If the substrate is cubic but terminated with a (110) surface, rhombohedral (001) or orthorhombic (001)
the epitaxial film would adopt a triclinic structure.\cite{Rondinelli:2012} Using the appropriate tables and the selection
rules, a polarized Raman experiment would be able to determine the mode and nature of substrate induced
strain effects on the BiFeO$_3$ epitaxial films.

\subsection{Limitations}
We believe we would be remiss if we omitted a discussion of some
limitations of the current paper and the methods used to derive the normal
mode distributions.
The normal modes in our paper were derived for simple, single phase, $AB$O$_{3}$
perovskites with octahedral rotations.
In mixed $A$- and $B$-site systems, e.g.\
La$_{1-y}$Sr$_{y}$Mn$_{1-x}M_{x}$FeO$_{3}$
($M$ = Cr, Co, Cu, Zn, Sc, or Ga),
the normal modes may be drastically different from their simple $AB$O$_{3}$
counterpart~\cite{Dubroka:2006}.
In these cases the $A$ and $B$ site disorder suppresses
the translational invariance that is a requirement for the range
of applicability of the symmetry analysis.
The result being that the lineshapes of the modes with atomic
displacement patterns from those sites could be
broadened.
Similar broadening could occur even in single phase perovskites grown in
thin film form, where  unintentional disorder and variability in the site occupation is induced
preferentially during growth along various directions, i.e., in the plane of or perpendicular to the substrate.
Such broadening can further lead to linewidths in which distinct modes of different,
but closely spaced, energies are indistinguishable.
Second, our results were derived for \emph{bulk} perovskites.
Their direct application to epitaxial films and superlattices may not be
immediately transferable.
As mentioned earlier, the Glazer octahedral tilt patterns are derived by considering
the octahedra as rigid units. Even though we allowed for the distortions of the $B$O$_6$ octahedra
to account for steric constraints, in epitaxial films and superlattices the octahedra are necessarily
further distorted.
A consequence of such geometrically required distortions is that they
can produce crystal structures with space groups different than those analyzed here, i.e.,
those obtained from Glazer's analysis of possible combinations of in- and out-of-phase
rotations.
Nevertheless, our schemes would be a good starting point in analyzing
these systems and deviations from ideality.
For example, a comparison based on the Raman mode predictions produced with
our tables to the actual experimental results could help  in
understanding the perturbative nature of octahedral engineering in
artificial perovskites.
Lastly, our treatment omits the consequences of magnetic ordering
on the crystal symmetry, which is expected to become critical in
perovskite superlattices formed with transition metal $B$-sites with open
$d$-shell configurations.
Several $AB$O$_{3}$ type perovskites undergo
paramagnetic metal to antiferromagnetic insulator transitions
followed by charge ordering as they are cooled through their N\'{e}el temperature.
The crystal symmetry operations in the antiferromagnetic state must
leave invariant not only the positions of the ions but also their magnetic
moments.
For example, magnetic ordering in the rutile structure reduces the space
group symmetry from that of rutile with non-magnetic cations ($P4_2/mnm$)
to $Pnnm$ or $D_{2h}^{12}$ \cite{Fleury:1968},
which in turn alters the Raman mode distribution.
We hope our paper spawns additional developments in the
experimental Raman studies of artificial perovskite oxides, theoretical approaches to
predict the normal mode distribution in such materials, and means to overcome
the current limitations discussed.
\section{Conclusions}
Using the nuclear site group analysis method we have
determined the complete set of normal modes and
the associated selection rules for all 15 of the $AB$X$_{3}$
perovskites systems with octahedral rotations (the Glazer systems).
For each mode, we have produced a corresponding
schematic diagram showing the vector displacement pattern
of the atoms participating in the particular mode.
The results of this analysis is
a compendium of 705 schematic diagrams, which
are now web accessible.

We have shown how some recent experimental findings can be analyzed using our
tables and the schematic diagrams to understand macroscopic interactions in the complex systems.
We expect these tables and the schematic diagrams to be useful tools
in analyzing the phase transitions, electron--lattice coupling and
elucidating structure--property relationships in $AB$O$_{3}$ perovskites
with octahedral rotations.
Equipped with this database for the bulk perovskites, we suggest that
one would be able to both analyze and rationally design functional artificial
materials built out of these perovskite blocks and rotated octahedral units.

\begin{acknowledgments}
Work supported by the ONR (N00014-11-1-0664).
J.M.R.\ was supported by ARO (W911NF-12-1-0133).
J.E.S.\ acknowledges support from the ARO (W911NF-08-1-0067).
Valuable discussions with Steven J.\ May are thankfully acknowledged.  The authors are also grateful to Pierre--Eymeric Janolin for helpful comments.
\end{acknowledgments}

%

\appendix
\begin{table*}[t]
\begin{ruledtabular}
\caption{System 1---Normal Modes for the Glazer system $a^0a^0a^0$.}
\begin{tabular}{lcccccc}
	   & $A$ & $B$ & $O$ &\multicolumn{3}{c}{Selection Rules} \\ [-0.1em]
\cline{5-7}
$a^0a^0a^0$ & $O_h$(1) & $O_h$(1) & $D_{4h}$(3) &  & \\
\hline
$T_{1u}$  		& 1	&	1	&	2	& $T$ & 	\\
$T_{2u}$  		& 	&		& 1 &  &  \\
\hline
Sum & 3	&	3	& 9 & & 15 & \\

\end{tabular}
\end{ruledtabular}
\end{table*}

\begin{table*}[t]
\begin{ruledtabular}
\caption{System 2---Normal Modes for the Glazer system $a^{0}a^{0}c^{+}$.}
\begin{tabular}{lccccccc}
	   & $A$ & $B$ & $O$(1) &$O$(2) &\multicolumn{3}{c}{Selection Rules}\\ [-0.1em]
\cline{6-8}
$a^{0}a^{0}c^{+}$ & $D_{2h}'$(2) & $C_{4h}$ (2) & $C_{4h}$(2) & $C_{2v}'$(4) & \\
\hline
$A_{1g}$ &  		& 	& 	&	1	&         & &$\alpha_{xx}+\alpha_{yy}, \alpha_{zz}$ 	\\
$A_{1u}$ &  		& 1	& 1 &	 	&         & &                                   	\\
$A_{2g}$ &  		& 	& 	&	1	& $R_{z}$ & & 	\\
$A_{2u}$ &  	1	& 1	& 1	&	1	& $T_{z}$ & &	\\
$B_{1g}$ &  		& 	& 	&	1	&         & &$\alpha_{xx}-\alpha_{yy}$ 	\\
$B_{1u}$ &  	1	& 	& 	&	1	&         & &	\\
$B_{2g}$ &  		& 	& 	&	1	&         & &$\alpha_{xy}$ 	\\
$B_{2u}$ &  		& 	& 	&		&         & & 	\\
$E_g$ &  		& 	& 	&	1	&  $(R_{x}, R_{y}$)& &  $\alpha_{xz}, \alpha_{yz}$ 	\\
$E_u$ &  	2	& 2	& 2	&	2	&  $(T_{x}, T_{y}$) & & 	\\
\hline
Sum & 6	&	6	& 6 & 12& &30 & \\
\end{tabular}
\end{ruledtabular}
\end{table*}

\begin{table*}[t]
\begin{ruledtabular}
\caption{System 3---Normal Modes for the Glazer system $a^0b^+b^+$.}
\begin{tabular}{lccccccccc}
	   & $A$(1) & $A$(2) & $A$(3) & $B$ & $O$(1) & $O$(2) &\multicolumn{3}{c}{Selection Rules}\\ [-0.1em]
\cline{8-10}
$a^0b^+b^+$ & $D_{2h}$ (4) & $D_{4h}$(2) & $D_{4h}$(2) & $C_{2h}'$(8) & $C_{s}^v$(16) & $C_{2v}'$(8)  \\
\hline
$A_{1g}$&  		& 	& 	&		& 2    & 1 &         & &$\alpha_{xx}+\alpha_{yy}, \alpha_{zz}$ 	\\
$A_{1u}$&  		& 	&  &	1 	& 1    &   &         & &                                 	\\
$A_{2g}$&  		& 	& 	&		& 1    & 1  & $R_{z}$ & &  	\\
$A_{2u}$&  	1	& 1	& 1	&	2	& 2    & 1 & $T_{z}$ & &	\\
$B_{1g}$&  		& 	& 	&		&  2   & 1 &         &  &$\alpha_{xx}-\alpha_{yy}$  	\\
$B_{1u}$&  		& 	& 	&	2	& 1    & 1 &	     &  &  \\
$B_{2g}$&  		& 	& 	&		& 1    & 1 &         & &$\alpha_{xy}$ 	\\
$B_{2u}$&  1	& 	& 	&	1	& 2    &   &         & &	\\
$E_g $&  		& 	& 	&		& 3    & 1 &  ($R_{x}, R_{y}$)& &  ($\alpha_{xz}, \alpha_{yz}$) 	\\
$E_u $&  	2	& 1	& 1	&	3	& 3    & 2 & ($T_{x}, T_{y}$) & & 	\\
\hline
Sum & 6	&	3	& 3 & 12 & 24 & 12 & & 60 & \\

\end{tabular}
\end{ruledtabular}
\end{table*}

\begin{table*}[t]
\begin{ruledtabular}
\caption{System 4---Normal Modes for the Glazer system $a^+a^+a^+$.}
\begin{tabular}{lccccccc}
	   & $A$(1) & $A$(2) & $B$ &$O$ & \multicolumn{3}{c}{Selection Rules}\\ [-0.1em]
\cline{6-8}
$a^+a^+a^+$ & $T_{h}$(2) & $D_{2h}$(6) & $S_{6}$(8) & $C_{s}$(24) &  & &   \\
\hline
$A_{g}$&  		& 	& 	&	2	&   &  &$\alpha_{xx}+\alpha_{yy}+\alpha_{zz} 	$\\
$A_{u}$&  		& 	&  1 &	1 	&    &   &                                    	\\
$E_{g}$&  		& 	&  	&	2	&      &   &    ($\alpha_{xx}+\alpha_{yy} -2 \alpha_{zz},
  \sqrt{3}\alpha_{xx}- \sqrt{3}\alpha_{yy}) $	\\
$E_u$&  		& 	& 1	&	1	&     & &	\\
$T_g$&  		& 	& 	&	4	& $R$  &  &$\alpha_{xx}, \alpha_{xz}, \alpha_{yz} $   	\\
$T_u$&  	1	& 3	& 3	&	5	 &$T$  &  &  \\
\hline
Sum & 3	&	9	& 12 & 36 & &  60 & \\

\end{tabular}
\end{ruledtabular}
\end{table*}

\begin{table*}[t]
\begin{ruledtabular}
\caption{System 5---Normal Modes for the Glazer system $a^+b^+c^+$.}
\begin{tabular}{lccccccccccc}
	   & $A$(1) & $A$(2) & $A$(3) &$A$(4)& $B$ & $O$(1) &$O$(2) & $O$(3) &\multicolumn{3}{c}{Selection Rules}\\ [-0.1em]
\cline{10-12}
$a^+b^+c^+$ & $D_{2h}$(2) & $D_{2h}$(2) & $D_{2h}$(2) &  $D_{2h}$(2)& $C_i$(8) & $C_{s}^{yz}$(8)& $C_{s}^{xy}$(8)& $C_{s}^{xz}$(8)   \\
\hline
$A_g$ &  		& 	& 	&		&      & 2 & 2 & 2 &   &   & $\alpha_{xx},\alpha_{yy}, \alpha_{zz} $	\\
$A_u$ &  		& 	&   &		& 3    & 1 & 1 & 1 &   &                              	\\
$B_{1g}$&  		& 	& 	&		&      & 1 & 2 & 1 & $R_{z}$ &  &$ \alpha_{xy} $	\\
$B_{1u}$&  	1	& 1	& 1	&	1	& 3    & 2 & 1 & 2 & $T_{z}$ &  &	\\
$B_{2g}$&  		& 	& 	&		&      & 1 & 1 & 2 & $R_{y}$ &   &$\alpha_{xz}  $	\\
$B_{2u}$&  	1	& 1	& 1	&	1	& 3    & 2 & 2 & 1 & $T_{y}$      & \\
$B_{3g}$&  		& 	& 	&		&      & 2 & 1 & 1 & $R_{x}$ &       &$\alpha_{yz}$ 	\\
$B_{3u}$&  1	& 1	& 1	&	1	& 3    & 1 & 2 & 2 & $T_{x}$ 	      &\\
\hline
Sum & 3	&	3	& 3 & 3 & 12 & 12 &12 & 12 & & 60 & \\

\end{tabular}
\end{ruledtabular}
\end{table*}

\begin{table*}[t]
\begin{ruledtabular}
\caption{System 6---Normal Modes for the Glazer system $a^0a^0c^-$.}
\begin{tabular}{lccccccc}
	   & $A$ & $B$ & $O$(1) &$O$(2) &\multicolumn{3}{c}{Selection Rules}\\ [-0.1em]
\cline{6-8}
$a^0a^0c^-$ & $D_{2d}'$(4) & $C_{4h}$(4) & $C_{2v}'$(8) & $D_4$(4)& & &  \\
\hline
$A_{1g}$&  		& 	& 1	&		&               &        &$\alpha_{xx}+\alpha_{xx}, \alpha_{zz} $	\\
$A_{1u}$&  		& 1	&   &		&               &           &                              	\\
$A_{2g}$&  		& 	& 1	&	1	&       $R_{z}$ &    &  	\\
$A_{2u}$&  	1	& 1	& 1	&	1	&       $T_{z}$ &    &	   \\
$B_{1g}$&  		& 	& 1	&		&               &    &    $\alpha_{xx}- \alpha_{yy}$  	\\
$B_{1u}$&  		& 	& 1	&		&               &           & \\
$B_{2g}$&  	1	& 	& 1	&		&       &      &$\alpha_{xy} $	\\
$B_{2u}$&   	& 	& 	&		&       &  & \\
$E_g$&  	1	& 	& 1 &	1	&      ($R_{x}, R_{y}$)&  & ($\alpha_{xz},\alpha_{yz}$) 	\\
$E_u$&  	1	& 2	& 2	&	1	&     ($T_{x}, T_{y}$) &	&    \\
\hline
Sum & 6	&	6	& 12 & 6  &    & 30  & \\

\end{tabular}
\end{ruledtabular}
\end{table*}

\begin{table*}[t]
\begin{ruledtabular}
\caption{System 7---Normal Modes for the Glazer system $a^0b^-b^-$.}
\begin{tabular}{lccccccc}
	   & $A$ & $B$ & $O$(1) &$O$(2) &\multicolumn{3}{c}{Selection Rules}\\ [-0.1em]
\cline{6-8}
$a^0b^-b^-$ & $C_{2v}^{z}$(4) & $C_{2h}^{x}$(4) & $C_{2}^{y}$(8) & $C_{2v}^{z}$(4)& & &  \\
\hline
$A_g$&  	1	& 	& 1	&	1	&            &    &             $\alpha_{xx}, \alpha_{yy}, \alpha_{zz}$ 	\\
$A_u$&  		& 1	& 1 &		&                 &                              	\\
$B_{1g}$&  		& 	& 2	&		&   $R_{z}$  &    &             $\alpha_{xy} $ 	\\
$B_{1u}$&  	1	& 2	& 2	&	1	&   $T_{z}$  &    &                             \\
$B_{2g}$&  	1	& 	& 1	&	1	&   $R_{y}$ &     &              $ \alpha_{xz}$ 	\\
$B_{2u}$&   1	& 2	& 1	&	1	&   $T_{z}$  &    &                               \\
$B_{3g}$&  	1	& 	& 2	&	1	&   $R_{x}$ &     &            $ \alpha_{yz} 	$\\
$B_{3u}$&   1	& 1	& 2	&	1	&   $T_{x}$ &	  &                            \\
\hline
Sum & 6	&	6	& 12 & 6  &  & 30 & \\

\end{tabular}
\end{ruledtabular}
\end{table*}

\begin{table*}[t]
\begin{ruledtabular}
\caption{System 8---Normal Modes for the Glazer system $a^-a^-a^-$.}
\begin{tabular}{lcccccc}
	   & $A$ & $B$ &  $O$ &\multicolumn{3}{c}{Selection Rules}\\ [-0.1em]
\cline{5-7}
$a^-a^-a^-$ & $D_3$(2) & $S_6$(2) & $C_2$(6) & & &\\
\hline
$A_{1g}$&  		& 	& 1	&	            &    &             $\alpha_{xx} , \alpha_{yy}, \alpha_{zz}$\\
$A_{1u}$&  		& 1	& 1 &	            &     &                         	\\
$A_{2g}$&  	1	& 	& 2	&	   $R_z$  &    &             	\\
$A_{2u}$&  	1	& 1	& 2	&	   $T_z$  &         &                             \\
$E_g$&  	1	& 	& 3	&	   $(R_x, R_y) $  &     & ($\alpha_{xx} - \alpha_{yy}, \alpha_{xy}), (\alpha_{xz} ,\alpha_{yz}$)\\
$E_u$&   1	& 2	& 1	&	   $T_x, T_y$   &     &                               \\
\hline
Sum & 6	&	6	& 18 &   & 30  & \\

\end{tabular}
\end{ruledtabular}
\end{table*}

\begin{table*}[t]
\begin{ruledtabular}
\caption{System 9---Normal Modes for the Glazer system $a^0b^-c^-$.}
\begin{tabular}{lcccccccc}
	   & $A_1$  & $B$ & $O$(1) & $O$(2) & $O$(3) &\multicolumn{3}{c}{Selection Rules}\\ [-0.1em]
\cline{7-9}
$ a^0b^-c^-$ & $C_s$(4) & $C_i$(4) & $C_2$(4) & $C_2$(4) & $C_s$(4) & &  & \\
\hline
$A_g$&  	2	& 	& 1	&	1	& 2    &  $R_z$       & &$\alpha_{xx}, \alpha_{yy}, \alpha_{zz}, \alpha_{xy} $ 	\\
$A_u$&  	1	& 3	& 1 &	1 	& 1    &  $T_z$       & &                                 	\\
$B_g$&  	1	& 	& 2	&	2	&  1   & $R_x, R_y$ & &$\alpha_{xx}, \alpha_{yy} $	\\
$B_u$&  	2	& 3	& 2	&	2	& 2    & $T_x, T_y$ & &	\\

\hline
Sum & 6	&	6	& 6 & 6 & 6  & & 30 & \\

\end{tabular}
\end{ruledtabular}
\end{table*}

\begin{table*}[t]
\begin{ruledtabular}
\caption{System 10---Normal Modes for the Glazer system $a^-b^-b^-$.}
\begin{tabular}{lcccccccc}
	   & $A_1$  & $B$ & $O$(1) & $O$(2) &\multicolumn{3}{c}{Selection Rules}\\ [-0.1em]
\cline{6-8}
$a^-b^-b^-$ & $C_2$(4) & $C_i$(4) & $C_1$(8) & $C_2$(4)  & &  & \\
\hline
$A_{g}$&  	1	& 	& 3	&	1	&   $R_{z}$       & &$\alpha_{xx}, \alpha_{yy}, \alpha_{zz}, \alpha_{xy}$  	\\
$A_{u}$&  	1	& 3	& 3 &	1 	&   $T_{z}$       & &                                 	\\
$B_{g}$&  	2	& 	& 3	&	2	&   $R_{x}, R_{y}$ & &$ \alpha_{xx}, \alpha_{yy}$ 	\\
$B_{u}$&  	2	& 3	& 3 &	2	&  $T_{x}, T_{y}$ & &	\\

\hline
Sum & 6	&	6	& 12 & 6 &   & 30 & & \\

\end{tabular}
\end{ruledtabular}
\end{table*}

\begin{table*}[t]
\begin{ruledtabular}
\caption{System 11---Normal Modes for the Glazer system $a^-b^-c^-$.}
\begin{tabular}{lccccccccc}
	   & $A$  & $B$(1) & $B$(2) & $O$(1)& $O$(2) & $O$(2) &\multicolumn{3}{c}{Selection Rules}\\ [-0.1em]
\cline{8-10}
$a^-b^-c^-$ & $C_1$ (2) & $C_i$(1) & $C_i$(1) & $C_1$(2) & $C_1$(2) & $C_1$(2) &  &  & \\
\hline
$A_g$&  	3	& 	&   & 3	&	3	& 3    &  $R$       & &$\alpha $ 	\\
$A_u$&  	3	& 3	& 3 &	3 	& 3   & 3 &  $T$       & &                                 	\\
\hline
Sum    &    6   & 3	& 3 &  6   &  6    &   6   & & 30 & \\

\end{tabular}
\end{ruledtabular}
\end{table*}

\begin{table*}[t]
\begin{ruledtabular}
\caption{System 12---Normal Modes for the Glazer system $a^0b^+c^-$.}
\begin{tabular}{lccccccccc}
	   & $A(1)$ & $A(2)$ & $B$  & $O(1)$& $O(2)$ & $O(3)$ &\multicolumn{3}{c}{Selection Rules}\\ [-0.1em]
\cline{8-10}
  $a^0b^+c^-$     & $C_{2v}^y$(4)  & $C_{2v}^y$(4) & $C_i$(8) & $C_{s}^{yz}$(8) & $C_{s}^{xy}$(8) & $C_{2}^x$(8)  \\
\hline
  $A_{g}$ &  	1	& 1	& 	&	2	& 2     & 1  &         &   &$\alpha_{xx},\alpha_{yy},\alpha_{zz} $	\\
  $A_{u}$ &  		& 	& 3 &	1	& 1    &  1  &         &   &                          	\\
  $B_{1g}$&  	1	& 1	& 	&	1	& 2    &  2  & $R_{z}$ &   & $\alpha_{xy}$ 	\\
  $B_{1u}$&  	1	& 1	& 3	&	2	& 1    &  2  & $T_{z}$ &   &	\\
  $B_{2g}$&  		& 	& 	&	1	& 1    &  2  & $R_{y}$ &   &$\alpha_{xz}$  	\\
  $B_{2u}$&  	1	& 1	& 3	&	2	& 2    &  2  & $T_{y}$ &    &\\
  $B_{3g}$&  	1	& 1	& 	&	2	& 1    &  1  & $R_{x}$ &       &$  \alpha_{yz}$ 	\\
  $B_{3u}$&   1	& 1	& 3	&	1	& 2    &  1  & $T_{x}$ &	      &   \\
\hline
Sum & 6	&	6	& 12 & 12 & 12  & 12 & & 60 & \\

\end{tabular}
\end{ruledtabular}
\end{table*}

\begin{table*}[t]
\begin{ruledtabular}
\caption{System 13---Normal Modes for the Glazer system $a^+b^-b^-$.}
\begin{tabular}{lccccccc}
	   & $A$ & $B$  & $O(1)$& $O(2)$ &\multicolumn{3}{c}{Selection Rules}\\ [-0.1em]
\cline{6-8}
$a^+b^-b^-$     & $C_{s}^{yz}$(4)  & $C_i$(4) & $C_{s}^{yz}$(4) & $C_1$(8) & &  \\
\hline
$A_{g}$ &  	2	& 	& 2	&	3	  &         &   &$\alpha_{xx},\alpha_{yy},\alpha_{zz} 	$\\
$A_{u}$ &  	1	& 3	& 1 &	3	  &         &   &                          	\\
$B_{1g}$&  	1	& 	& 1	&	3	  & $R_{z}$ &   &$ \alpha_{xy} 	$\\
$B_{1u}$&  	2	& 3	& 2	&	3	  & $T_{z}$ &   &	\\
$B_{2g}$&  	2	& 	& 2	&	3	  & $R_{y}$ &   &$\alpha_{xz}  $	\\
$B_{2u}$&  	1	& 3	& 1	&	3	  & $T_{y}$ &    &\\
$B_{3g}$&  	1	& 	& 1	&	3	  & $R_{x}$ &       &$  \alpha_{yz} $	\\
$B_{3u}$&   2	& 3	& 2	&	3	  & $T_{x}$ &	      &   \\
\hline
Sum & 12	&	12	& 12 & 24  & & 60 & \\

\end{tabular}
\end{ruledtabular}
\end{table*}

\begin{table*}[t]
\begin{ruledtabular}
\caption{System 14---Normal Modes for the Glazer system $a^+b^-c^-$.}
\begin{tabular}{lccccccccccc}
	   & $A$(1)  & $A$(2) & $B$(1) & $B$(2) & $O$(1) & $O$(2)& $O$(3)& $O$(4) & \multicolumn{3}{c}{Selection Rules}\\ [-0.1em]
\cline{10-12}
$ a^+b^-c^-$ & $C_s$(2) & $C_s$(2) & $C_i$(2) & $C_i$(2) & $C_1$(4) & $C_s$(2)  & $C_s$(2)& $C_1$(4) & & & \\
\hline
$A_{g}$&  	2	& 2	& 	&		& 3    &  2 & 2 & 3 & $R_{z}$       & &$\alpha_{xx}, \alpha_{yy}, \alpha_{zz}, \alpha_{xy}  $	\\
$A_{u}$&  	1	& 1	& 3 &	3 	& 3    & 1 & 1 & 3 & $T_{z}$       & &                                 	\\
$B_{g}$&  	1	& 1	& 	&		& 3   & 1 & 1 & 3 &$R_{x}, R_{y}$ & & $\alpha_{xx}, \alpha_{yy}$ 	\\
$B_{u}$&  	2	& 2	& 3	&	3	& 3    & 2 & 2 & 3 & $T_{x}, T_{y}$ & &	\\

\hline
Sum & 6	&	6	& 6 & 6 & 12 & 6 & 6 & 12 & & 60 & \\

\end{tabular}
\end{ruledtabular}
\end{table*}

\begin{table*}[t]
\begin{ruledtabular}
\caption{System 15---Normal Modes for the Glazer system $a^+a^+c^-$.}
\begin{tabular}{lcccccccccc}
	   & $A(1)$ & $A(2)$ & $A(3)$ & $B$ & $O(1)$& $O(2)$ & $O(3)$ &\multicolumn{3}{c}{Selection Rules}\\ [-0.1em]
\cline{9-11}
$a^+a^+c^-$ & $C_{2v}^4$(4) & $D_{2d}$(2) & $D_{2d}$(2) & $C_i$(8) & $C_{s}^v$(8) & $C_v$(8)& $C_{2}'$(8) & & &  \\
\hline
$A_{1g}$&  1    & 	& 	&		& 2    & 2 & 1 &       &    &$\alpha_{xx}+\alpha_{yy}, \alpha_{zz}$ 	\\
$A_{1u}$&  		& 	&   &	3 	& 1    & 1 & 1 &       &                                 	\\
$A_{2g}$&  		& 	& 	&		& 1    & 1 & 2 &$R_z$ & &  	\\
$A_{2u}$&  	1	& 1	& 1	&	3	& 2    & 2 & 2  &$T_z$ & &	\\
$B_{1g}$&  	1	& 1	& 1	&		& 2    & 2 & 2  &    &    &$\alpha_{xx}-\alpha_{xy}  $	\\
$B_{1u}$&  		& 	& 	&	3	& 1    & 1 & 2 	&    &  \\
$B_{2g}$&       & 	& 	&  	    & 1    & 1 & 1 &   & &$\alpha_{xy} $	\\
$B_{2u}$&  1	& 	& 	&	3	& 2    & 2 & 1 &     &	\\
  $E_g$&   2	& 1	& 1	&		& 3    & 3 & 3 &($R_x, R_y$)&   &  ($\alpha_{xz}, \alpha_{yz}$) 	\\
$E_u$&   2	& 1	& 1	&	6	& 3    & 3 & 3 & ($T_x, T_y$) & & 	\\
\hline
Sum & 12	&	6	& 6 & 24 & 24 & 24 & 24 & & 120 & \\

\end{tabular}
\end{ruledtabular}
\end{table*}

\begin{table*}[t]
\begin{ruledtabular}
\caption{System 16---Normal Modes for the Non Glazer system BiFeO$_3$ $a^-a^-a^-$.}
\begin{tabular}{lcccccc}
	   & $B$ & $Fe$ & $O$ &\multicolumn{3}{c}{Selection Rules} \\ [-0.1em]
\cline{5-7}
$a^-a^-a^-$ & $C_3$(2) & $C_3$(2) & $C_{1}$(6) &   \\
\hline
$A_{1}$ & 1	&	1	& 3	& $T_z$&&$\alpha_{xx}^{z}+\alpha_{yy}^{z}, \alpha_{zz}^{z}$  	\\
$A_{2}$ & 1	&	1	& 3 &  $R_z$  \\
$E$     & 2	&	2	& 6 &  ($T_x$, $T_y$);($R_x$$R_y$)&&$(\alpha_{xx}^{y}-\alpha_{yy}^{y}, \alpha_{xy}^{x}), (\alpha_{xz}^{x}, \alpha_{yz}^{y})$ \\
\hline
Sum & 6	&	6	& 18 & & 30 & \\

\end{tabular}
\end{ruledtabular}
\end{table*}

\end{document}